\documentclass[preprint,prd,nofootinbib]{revtex4}

\usepackage{color} 
     
\usepackage{amssymb}
\usepackage{amsmath}
\usepackage{graphicx}
\begin{document}

%
%

\title{Matter-antimatter asymmetry  and other cosmological puzzles via running vacuum cosmologies}
\author{J. A. S. Lima}
\email{jas.lima@iag.usp.br}
\affiliation{Departamento de Astronomia, Instituto de Astronomia Geof\'isica e Ci\^encias Atmosf\'ericas, Rua do Mat\~ao 1226, 05508-900, S\~ao Paulo, SP, Brazil }


\author{D. Singleton}
\email{dougs@mail.fresnostate.edu}
\affiliation{Department of Physics, California State University, Fresno, Fresno, CA 93740-8031, USA}





\begin{abstract}
Current astronomical observations are successfully explained by the present cosmological paradigm  based on the concordance model ($\Lambda_0$CDM + Inflation). However, such a scenario  is composed of a heterogeneous mix of ingredients for describing the different stages of cosmological evolution.   Particularly,  it does not give an unified explanation connecting the early and late time accelerating inflationary regimes which are separated by many aeons. Other challenges to the concordance model include: a singularity at early times or the emergence of the Universe from the quantum gravity regime,  the ``graceful'' exit from inflation to the standard radiation phase, as well as, the  coincidence and cosmological constant problems. We show here that a simple running vacuum model or a time-dependent vacuum may provide insight to some of the above open questions (including a complete cosmic history), and also can explain the observed matter-antimatter asymmetry just after the initial deflationary period.
\end{abstract}

\maketitle



\section{Introduction}	

The observed amount of antimatter is relatively small being usually explained by a secondary production in high energy cosmic ray collisions or in catastrophic astrophysical events. In this way,  one the oldest puzzles in cosmology can be put colloquially as ``Why is there more stuff than anti-stuff"; or more technically why is there more matter than anti-matter; or even more specifically why are there more baryons than anti-baryons. This puzzle occurs because the fundamental interactions of particle physics appear (at first glance) to be symmetric with respect to how matter and anti-matter interact. Thus, the simply proposing that the Universe started off with more matter than anti-matter is unsatisfying unless one accepts the anthropic principle {\it i.e.} ``if this were not the case then we would not be here to think about this". 

In the late 1960s Andrei Sakharov gave some general {requirements\cite{Sakharov}} for naturally generating a matter-antimatter asymmetry. The three conditions are: (i) Baryon number must be violated by some of the fundamental interactions; (ii) The interactions must violate C-symmetry and CP-symmetry; (iii) The interactions must operate for sometime out of thermal equilibrium. This general recipe was very useful to implement different solutions to the baryogenesis problem. However,  it is now widely believed that such a prescription though very natural does not need to be strictly {adopted\cite{Rev1,Rev2,Baryon}} so that many different approaches have been suggested in the {literature\cite{Dine03,Stein2004,Rev2016,LS2016,OSP2017, LS2}}. 

On the other hand, current astronomical observations are well described by the new cosmological paradigm based on the concordance model ($\Lambda_0$CDM + Inflation). Nevertheless,  in order to explain the different stages of the cosmic evolution: inflation, radiation, dark matter and dark energy,  the model is composed by a mixture of many different ingredients. In particular,  it does not provide a unified explanation connecting smoothly the early and late time accelerating inflationary regimes which are separated by many aeons. Other challenges to the concordance model {include\cite{Wein89,Pad03,Peeb2003,Rev2004}}: a singularity at early times or the emergence of the Universe from the quantum gravity regime,  the ``graceful'' exit from inflation, the  coincidence and cosmological constant problems \footnote{There are also the so-called  small scale puzzles, like the``missing satellite problem", that is, the discrepancy between the number of predicted subhalos in N-body simulations and the observed one, as well as,  the ``Cusp/Core" and the ``Angular-Momentum Catastrophe". However, such problems are heavily dependent on the local astrophysical processes (see, for instance,  Del Popolo et al.\cite{DP2014} and references there in for a recent discussion).}.

In this article,  we investigate a class of nonsingular cosmological models describing the early and late time cosmic evolution (complete cosmological history) which is also able to generate the baryon asymmetry observed in the Universe. The models are powered by a running vacuum cosmology (or time-dependent cosmological constant model) and start their  evolution  from a deflationary Universe (de Sitter phase) and evolves to the current cosmic concordance  $\Lambda_0$CDM cosmology.

\section{Running Vacuum Cosmologies}

Running vacuum cosmologies can be justified both phenomenologically and from first principles based on quantum field theory (QFT) in curved {spacetimes\cite{OT86,PRD92,Waga93,PRD94,PRD96}}. The vacuum state of all existing fields can be represented by an energy-momentum tensor (EMT) which reflects the Lorentz invariance of its energy density and pressure.  By averaging over all fields the minimally coupled EMT in general relativity  reads  $\langle T_{\mu \nu} \rangle  = \langle \rho \rangle g_{\mu \nu}$. Transferring the geometric cosmological constant $\lambda$ to the {\it r.h.s.} of Einstein's equations leads to an effective cosmological ``constant" term, $\Lambda = \lambda + \frac{\langle \rho \rangle }{M_{Pl}^{2}}$, where $M_{Pl}=(8\pi G)^{-1/2}\sim 2.4\times 10^{18}$ GeV is the reduced Planck {mass\cite{Wein89}}.

The field equations  with this effective $\Lambda$ can be written as:

\begin{equation}
G^{\mu \nu} = M_{Pl}^{-2}\,\large[\,{T^{\mu\nu}_{(f)}} + \Lambda M_{Pl}^{2}g^{\mu \nu}\,\large]\,,
\end{equation}
where the energy-momentum tensor of the fluid is given by:
\begin{equation}
{T^{\mu\nu}_{(f)}}=  (\rho + p)u^{\mu}u^{\nu} - p g^{\mu\nu}\,, 
\end{equation}
where $\rho$ and $p$ are the corresponding energy density and pressure.

Because the Einstein tensor is divergenceless ($\nabla_{\mu} G^{\mu \nu} = 0$), it is usually argued that the effective vacuum energy density remains constant ({\it i.e.} $\nabla_{\mu} \Lambda=\partial_{\mu} \Lambda \equiv 0$). However, such a result holds only if the EMT of the cosmic fluid is always separately conserved ($\nabla_{\mu} {T^{\mu\nu}_{(f)}}=0$). In reality, one finds that a  running $\Lambda$ can transfer energy to the cosmic fluid (${T^{\mu\nu}_{(f)}}$)    
\begin{equation}
\nabla_{\mu} G^{\mu \nu} = 0 \,\, \Rightarrow \nabla_{\mu}\, {T^{\mu\nu}_{(f)}} = - g^{\mu \nu}\nabla_{\mu} \Lambda\,,
\end{equation}
and this exchange of energy between both components also implies that particles of the fluid component can be produced from the decaying vacuum. For instance, in the flat Friedman-Robertson-Walker geometry adopted here ($c=1$):

\begin{equation}
ds^{2} = dt^{2} - a^{2}(t)\large [dx^{2} + dy^{2} +  dz^{2}\large]\,,
\end{equation}
where $a(t)$ is the scale factor, the energy conservation law for the total energy-momentum tensor reads:
\begin{equation}
\dot \rho +  3H(\rho + p) = -M_{Pl}^{2}{\dot \Lambda}\,,
\end{equation}
where $H = \dot a/a$ is the Hubble parameter. The balance  equation above shows that if $\dot \Lambda < 0$, energy is transferred  from the decaying vacuum to the fluid component. In addition, in the case of an ``adiabatic" decay, the comoving number of particles is also a function of the cosmic {time\cite{Lima96}}.

The assumption  of a running vacuum energy density also seems physically more appealing than the standard view of a {\it fixed} $\Lambda$-term. Due to the cosmological principle, the physical quantities at the background level should depend only on time and the condition $\dot \Lambda \neq 0$, means that the extremely low value currently observed for the vacuum energy density can be theoretically  accommodated. The present smallness of $\Lambda_0 \equiv \Lambda (t_0)$ can be seen as a natural consequence of the oldness of the Universe. Some consequences of this view has also an observational support. Recent works  have shown that a battery of tests including {\it SNIa, BAO, H(z), LSS, BBN, CMB} based on running vacuum models provide a fit better than the standard $\Lambda_0$CDM {cosmology\cite{AS2015,Sola2016}}.  

A basic question in running vacuum models is how does $\Lambda$ vary as the Universe evolves? Probably, the most successful path comes from QFT techniques based on the renormalization group (RG) approach in curved spacetimes. The vacuum energy density `runs' because the effective action inherits quantum effects from the matter sector. Generically, RG techniques in curved spacetimes lead to a dependence $\rho_\Lambda (H) = M_{Pl}^{2}\Lambda (H)$ defined by the {relation\cite{Sola2016,PLB000,PRD2013,LBS2013,LBS2015,LBS2016}}  
\begin{equation}
\label{RG-rho}
\frac{d\rho_{\Lambda }}{d \ln H^2} = \frac{1}{16 \pi^2} \sum _i \left[ a_i M_i ^2 H^2 + b_i H^4 + c_i \frac{H^6}{M_i ^2} +...\right]\,,
\end{equation}  
where H is the Hubble parameter, $a_i, b_i, c_i$ are dimensionless coefficients and $M_i$ are masses of the fields in the model which are taken to be the GUT scale or smaller. The reason for selecting even powers of H is because of the general covariance of the effective action of QFT in curved spacetimes. Integrating equation \eqref{RG-rho} leads to a power series of the running vacuum  $\Lambda (H) = c_0 + c_2H^2 + c_4 H^{4} + c_6H^{6} + ...$, where $c_0$ provides the dominant term at very low energies when $\Lambda (H_0)  = \Lambda_0$ and the other coefficients ($c_2, c_4, c_6...$) are in terms of $a_i, b_i, c_i, M_i, M_{Pl}$. The $H^{2}$ term is natural from dimensional grounds, but it is observationally very small because the observed Universe is $\Lambda_0CDM$-like to great accuracy. The $H^{4}$ term also arises from free fields in an earlier de Sitter {stage\cite{D77,Davies77,CT09}}. The weakness of the above RG approach in determining the running of the vacuum energy density is that there is no way to fix the renormalization coefficients $a_i, b_i, c_i$ which in turn means that $c_2, c_4, c_6...$ are also unfixed parameters. In essence one has parametrized the effect of the $H^{2n}$ terms, but one does not really have a definite prediction. Despite this weakness of the RG in curved spacetime approach to giving specific values for the renormalization coefficients one can still be confident in the general form of the running of $\rho_{\Lambda }$ as an expansion in powers of $H$, which is supported by the effective field theory approach to gravity as discussed long ago by {Donoghue\cite{donoghue}}, as well as to a some degree by $f(R)$ and $f(R,T)$ gravity {theories\cite{fR,Harko2011}} (see also Refs. in the quoted papers). 
 
Inspired by this approach we consider here a  specific parametrization of $\Lambda (H)$  with $c_0 = \Lambda _0$, $c_2 \approx 0$ (this choice for $c_2$ is supported by observations) and $c_4 = 3/H_I ^2$, where $H_I$ is an inflationary scale and the factor of 3 is only for mathematical convenience.  Thus, neglecting all higher power terms we adopt the following decaying $\Lambda$-term: 
\begin{equation}
\label{lh1}
\Lambda (H) = \Lambda_0 +  3\frac{H^{4}}{H_I^{2}}\,.
\end{equation}
The advantage of this parametrization of the running vacuum model  is that it evolves naturally to the current $\Lambda_0$CDM for $H \ll H_I$. The evolution equation for the Hubble parameter reads (see {references\cite{PRD2013,LBS2013,LBS2016}} for a more general case):
\begin{equation}
\label{HE}
\dot H+\frac{3}{2}(1+\omega)H^2\left[1-
\left(\frac{H}{H_I}\right)^{2}\right] - \frac{(1+\omega)}{2}{\Lambda_0}=0\,,
\end{equation}
where a dot means time derivative and $\omega \equiv {p}/{\rho}$ is the equation of state parameter (EoS). At early times, the $\Lambda_0$ term above is negligible and $\dot H = 0$ at $H=H_I$. This means that the Universe began its evolution from an unstable, non-singular de Sitter inflationary state. The analytical solution to \eqref{HE} for a radiation fluid ($\omega=\frac{1}{3}$) and $\Lambda_0 \ll H^{2}$ is:

\begin{equation}
\label{Ha-1}
H(a) = \frac{{H _I}}{\sqrt {1+  ({a}/{a_{end}})^{4}}}\,\,,
\end{equation}
where $a_{end}$ is  the scale factor when $\rho_{\Lambda}=\rho_{rad}$, which coincides with the end of inflation [\,${\ddot a}=0$ and $H(a_{end})\equiv H_{end}=H_I/{\sqrt 2}$\,].  Therefore, for $a \ll a_{end}$, we find $H=H_I$ as expected for an early de Sitter phase, while for $a \gg a_{end}$,  the model evolves smoothly to the standard radiation phase [\,$a(t) \propto t^{1/2}$\,]. After radiation-matter equality, the model becomes dominated by cold dark matter ($\omega =0$), and since $\Lambda_0$ is different from zero, the model approaches the current $\Lambda_0$CDM cosmology.

The minimal scenario outlined above is free of an initial singularity, and, thus, has no horizon problem. Its smooth transition from an early de Sitter to the radiation phase provides  a parametrization to a ``graceful'' exit from the initial inflationary stage. In this case, the temperature evolution and the increasing entropy can also be analytically determined. The entropy of the effectively massless particles is initially zero in the de Sitter stage,  but increases continuously reaching the equilibrium value in the radiation adiabatic phase ($\dot \Lambda =0$). It has been {proved\cite{LBS2015}} that such running vacuum cosmology also generates the amount of entropy within the current Hubble radius, $S_0 \sim 10^{88}$ (in natural units).

\section{Baryon asymmetry in a running vacuum model}

Let us now discuss the matter-antimatter asymmetry in the framework of this running vacuum cosmology. As we shall see,  the model provides an acceptable baryon asymmetry (B-asymmetry for short), and the same physical ingredient (running vacuum) may also shed some light on the coincidence and cosmological problems. 

The B-asymmetry is characterized  by the dimensionless $\eta$-parameter\cite{eta}:
\begin{equation}\label{limit}
5.7\times 10^{-10} < \eta  \equiv \frac{n_B - n_{\bar B}}{s}< 6.7 \times 10^{-10}\,,
\end{equation}
where $n_{B}$, $n_{\bar B}$ are  the number densities of baryons (anti)-baryons,  respectively, and $s$ is the present  day radiation entropy density. 

To generate the B-asymmetry we adopt a new approach based on the derivative coupling between the baryon current $J_B ^\mu$ and the running vacuum whose action {reads\cite{LS2}}: 
\begin{equation}
{\cal A}=\frac{1}{M_{\Lambda}^{2}}\int {d^4}x{\sqrt{-g}}(\partial_\mu\Lambda)J^{\mu}_B\,,
\end{equation}  
where $M_{\Lambda}$ is an unknown cut-off mass scale of the effective running vacuum theory. From this coupling  one can define an effective chemical potential for a species $i$ of baryons/anti-baryons and the resulting $\eta$ as:
\begin{equation}
\label{chem-pot-1}
\mu _i =  \pm \frac{{\dot\Lambda}}{M^2_{\Lambda}}\,\, \rightarrow \,\,\eta = \frac{n_B}{s} \approx -\frac{{\dot \Lambda}}{M_{\Lambda}^{2} T} {\Big \rvert} _{T=T_D}\,. 
\end{equation}
The $\pm$ in $\mu _i$ is the charge of the baryons/anti-baryons, respectively (particles have baryon/anti-baryon number of $\pm 1$). In the expression  for $\eta$ we have approximated, $15g_b/4\pi^{2}g_* \sim {\cal O} (1)$, where $g_b$ counts all possible baryonic  degrees of freedom and $g_*$  counts all the massless degrees of freedom. Note also that both $\dot \Lambda$ and the  temperature, $T$, must be evaluated when the B-violation operator decouples ({\it i.e.} when $H=H_D, T=T_D$).

The amount of B-asymmetry is determined once $T_D=T(H_D)$ and $\dot \Lambda (H_D)$ are known. In our  model $\dot \Lambda$ and the radiation temperature law (from $\rho_{r} \propto T_{r}^{4}$) read  \cite{LBS2015}  
\begin{equation}
\label{dotLT}
{\dot \Lambda } = -24H^3\left(\frac{H}{H_I}\right)^{2} 
\left[ 1- \frac{H^2}{H_I^2} \right] \,\,\,\,\,\,\, {\rm and} \,\,\,\,\,\,\,
T_{r}(H) = \sqrt2 T_{end}\left(\frac{H}{H_I}\right)^{1/2}\left[1 - \frac{H^2}{H_I ^2}\right]^{\frac{1}{4}}\,,
\end{equation}
where $T_{end}= T(H_{end})\simeq 10^{-\frac{1}{4}}\sqrt{M_{Pl}H_I}$ is the temperature at the end of inflation.

The primeval de Sitter stage ($H=H_I$) is a particle / anti-particle symmetric spacetime because $\eta\equiv 0$  since ${\dot \Lambda} (H_I) = 0$.  After the de Sitter stage $\eta$ initially increases as $H$ decreases. After  $\eta$  reaches its maximum value it asymptotically goes to zero again in the radiation FRW phase ($H \ll H_I$). Hence the $\eta$ value should be frozen to its observed value at some moment ({\it i.e.} at the decoupling time) before the radiation era is reached. Note also that initially there is no thermal bath [$T_{r} (H_I) \equiv 0$], but the temperature increases very fast due to  running {vacuum\cite{LBS2015}}. After its maximum value, the temperature decreases ($T_r\propto H^{1/2}$) as the Universe enters in the standard radiation phase ($H \ll H_I$). 

\vskip 0.3cm
\begin{table}[ph]
\caption{Baryogenesis prediction in running vacuum cosmologies. The values are in good agreement with the current baryonic $\eta$-constraints and the inflationary scale, $H_I$.}
{\begin{tabular}{@{}ccc@{}} \toprule
\,\,$M_{\Lambda}/M_{Pl}\,\leq 1$ & \,\,\, $H_I/M_{Pl}\,\leq 1$  &  $\eta$ \\ \toprule 
\hphantom{0}9.5 $\times 10^{-4}$ & \hphantom{0} 4.1 $\times 10^{-7}$  & \hphantom{0} 6.2 $\times 10^{-10}$\\  
1.8 $\times 10^{-3}$ & \hphantom{0} 6.7 $\times 10^{-7}$  & \hphantom{0} 5.9 $\times 10^{-10}$\\  
9.6 $\times 10^{-2}$ & \hphantom{0} 1.6 $\times 10^{-5}$  & \hphantom{0} 5.8 $\times 10^{-10}$\\  
5.7 $\times 10^{-2}$ & \hphantom{0} 1.1 $\times 10^{-5}$  & \hphantom{0} 6.4 $\times 10^{-10}$\\  
1.0                  & \hphantom{0} 1.1 $\times 10^{-4}$  & \hphantom{0} 6.6 $\times 10^{-10}$\\  \botrule 
\end{tabular}\label{ta1}}
\end{table}

The typical time scale of the B-asymmetry chemical potential from Eq. \eqref{chem-pot-1} is $\mu_{Bi}/\dot \mu_{Bi} = |\dot \Lambda/\ddot \Lambda|$, and, as such,  a quite accurate decoupling condition for RV models can be defined by $|\dot \Lambda/\ddot \Lambda|^{-1}H\simeq 1$. A lengthy but straightforward calculation provides the value of the Hubble parameter at the decoupling time \footnote{ The same result is also obtained via the decoupling of a 6-dimensional B-violating operator in a generic GUT, ${\cal L}_{\Delta B} = \frac{1}{M_X ^2} \epsilon ^{\alpha \beta \gamma} \epsilon^{ab} ({\bar u}^c _\gamma \gamma^\mu q_{\beta a}) ({\bar d}^c _\alpha \gamma_\mu l_b)$, where  $M_X\sim H_I$ is the mass scale,  and $u, d, q, l$ stand for up-quark, down-quark, quark and lepton respectively.},  $H_D = \left(\frac{9}{14}\right)^{1/2} H_I$, which can be inserted into \eqref{dotLT} to give ${\dot \Lambda (H_D)}$ and $T_D= T_{rad}(H_D)$. 
Inserting  these results into $\eta$ from \eqref{chem-pot-1} gives
\begin{equation}\label{etaD2}
\eta \approx  5.2\,\left(\frac{M_{\Lambda}}{M_{Pl}}\right)^{-2}\left(\frac{H_I}{M_{Pl}}\right)^{5/2}.
\end{equation}

In Table 1 we display the running vacuum predictions for selected values of the sub-Planckian dimensionless ratios, $M_{\Lambda}/M_{Pl}$ and $H_I/M_{Pl}$. These free parameters provide both the observed baryon  $\eta$-asymmetry and agreement with the inflationary {scale\cite{eta,LS2,planck}} $H_I$. In principle there is no fine-tuning because these two physical constraints are satisfied for different combinations of two free sub-Planckian parameters ($M_{\Lambda},\, H_I$). We also stress that the proposed baryogenesis mechanism does not require new ingredients since it is powered by the same entity (the running vacuum) driving the evolution of the spacetime. 

\section{Conclusions}
 
Summarizing, in this work we have discussed the generation of the baryon asymmetry  in the context of a  nonsingular  running vacuum model where the time-dependent 
$\Lambda_H$-term  evolves obeying the scaling law: $\Lambda (H) = \Lambda_0 + 3\frac{H^{4}}{H_I^{2}}$, where $H_I$ is the scale of the inflationary phase occurring in the early Universe. The model  describes the entire cosmological history, namely, from inflation to the current accelerating dark energy epoch  powered by the present day vacuum energy density ($\Lambda_0$).   In general lines, it can be seen as a kind of emergent Universe because it is nonsingular and behaves  in the infinity past ($t \rightarrow -\infty$) as a stationary spacetime (de Sitter phase). As should be expected, at late times it approaches the cosmic concordance model.  

As we have seen, the model is also able to generate the observed baryon asymmetry in the Universe. In addition, it also shed some light on several other cosmological puzzles: (i) avoiding the initial singularity (and therefore has no horizon problem); giving a smooth exit from inflation to the radiation phase; (iii) as shown in reference\cite{LBS2015} generating all the observed entropy from a primeval nonsingular de Sitter stage. Note also that for $H \ll H_I$, all successes of the $\Lambda_0$CDM cosmology are also preserved. The upshot is a unified vacuum picture, spanning the entire history of the universe (from an early de Sitter stage to a very late time de Sitter cosmology) thereby evolving naturally through the current $\Lambda_0$CDM cosmology. It is also interesting that given the high values of $H_I$ on Table 1, one may see that the ratio $\Lambda_I/\Lambda_0 \sim (H_I/H_0)^{2}$ is also in agreement with the so-called cosmological constant problem.

Finally, it is worth mentioning  that the new line of inquire  presented here for some open cosmological questions is not an exclusive characteristic of this phenomenological variable $\Lambda(H)$ models evolving from de Sitter to de Sitter. Actually, at the  the background level such models can be mimicked by ``adiabatic"particle creation cosmologies \cite{Prigogine89, CLW92}. This happens when the gravitationally induced  particle production rate $\Gamma (t) =  3H\Lambda(t)/\rho$. The dynamical equivalence  with the concordance model has already been discussed in a series of paper\cite{LJO2010,LBC2012,GCL2014,LL2014}. Actually, this reduction of the dark sector may simulate  the $\Lambda(t)$ model discussed here  both at the background and for constant $\Lambda_0$ also in  perturbative {level\cite{RSW2014a,RSW2014b,LSC2016}}. Studies for the general case are in progress, but such alternatives to vacuum decay cosmologies are by construction unable to shed some light on the cosmological constant problem.  

{\par\noindent {\bf Acknowledgments:}} JASL was partially supported by CNPq, CAPES (PROCAD 2013) and FAPESP (Brazilian Research Agencies).

\end{document}